\documentclass[10pt,preprint2]{aastex}

\usepackage{bm}

\shorttitle{Rotation Curve Asymmetry}
\shortauthors{Hodge}

\begin{document}

\title{Neighboring Galaxies' Influence on Rotation Curve Asymmetry}

\author{John C. Hodge\altaffilmark{1} and Michael W. Castelaz\altaffilmark{2}}
\affil{Pisgah Astronomical Research Institute, 1 PARI Drive, Rosman, NC, 28772}
\altaffiltext{1}{Visiting from XZD Corp., 3 Fairway St., Brevard, NC, 28712, scjh@citcom.net}
\altaffiltext{2}{mcastelaz@pari.edu}

\begin{abstract}

The rotation velocity asymmetry v observed in spiral galaxy HI rotation curves linearly correlates with the effective potential force from the 10 closest neighboring galaxies normalized for the test particle mass and the gravitational constant.  The magnitude of the potential force from a close galaxy is proportional to the luminosity of the close galaxy and inversely proportional to the square of the distance from the close galaxy to the target galaxy.  The correlation coefficient is 0.99 and F test is 0.99.  Also, the slope of the rotation curve in the disk region of a galaxy from rising to flat to declining is qualitatively correlated with increasing asymmetry and, hence, to the net force from other galaxies.  The result is based on a sample of nine spiral galaxies with published Cepheid distances and rotation curves and with a wide range of characteristics.  These relationships are interesting not only for their predictive power but also because (1) they suggest a galaxy's dynamics and the shape of its rotation curve are related to the potential energy and distance of neighboring galaxies, (2) they suggest a rising rotation curve in the disk region is intrinsic, (3) they suggest the Tully-Fisher relationship's assumption that the mass to intrinsic luminosity ratio is constant among galaxies is valid, and (4) they are inconsistent with MOND, dark matter, and the Linear Potential Model.

\end{abstract}
\maketitle
\section{INTRODUCTION}

Asymmetry is often observed in rotation curves of spiral galaxies.  Like the rotation curve, asymmetries indicate the forces and dynamics in a galaxy.  Small deviations in the rotation curves of a few $kms^{-1}$ are well known \citep{shane}.  Large deviations are less appreciated \citep{jog} because the observational data are generally averaged.  Hence, only highly asymmetric cases such as NGC 4321 \citep{knap} and NGC 3031 \citep{rots} are recognized.  The rotation curves of all the nearby galaxies where the kinematics are studied have asymmetry such as NGC 0224 \citep{simi} and NGC 0598 \citep{coli}.  Asymmetry has also been reported in the inner parts of the optical disk \citep{sofu}.  Rotation curve asymmetry appears to be the norm rather than the exception \citep{jog}.  \citet{wein}and \citet{jog97} proposed the implied mass asymmetry is due to a galaxy interaction.  In this Paper, the rotation curve asymmetry from our viewpoint is related to the strength of the forces from neighboring galaxies.  A qualitative correlation of rotation curve shape with rotation curve asymmetry is shown.  This result suggests the rotation curve shape is naturally rising and neighboring galaxies cause flat and declining rotation curves.  These results are inconsistent with MOND, dark matter, and the Linear Potential Model.

\section{MODEL}

\citet{hodg} (hereinafter Paper 1) found the observed magnitude of a target galaxy depends on the radiant energy of neighboring galaxies and on the intrinsic luminosity of the target galaxy.  The potential energy field of neighboring galaxies may also produce observable effects in target galaxies.

Consider a particle in the disk region of a spiral galaxy.  If there are no outside forces, the intrinsic orbit of the particle will be achieved by a delicate balance of the intrinsic forces of the galaxy.  The circular orbits of galaxy disks (\citet{binn}, pages 723-4) imply the intrinsic forces are centrally directed.  Therefore, a small, external force directed nearly uniformly across the plane of the galaxy may cause a perturbation in the orbit and in the rotational velocity of particles in the disk.  Thus, the asymmetry of the rotation velocity curve may be a sensitive indicator of the external forces exerted on a galaxy's particles.

To test the effects of the potential energy field of neighboring galaxies, a measure of galaxy mass is needed.  Commonly, mass is derived from rotation curve models.  However, such mass calculations are lacking for most galaxies.  Paper 1 and \citet{aaro} suggested the success of the Tully-Fisher relationship (TF) implies that the ratio of intrinsic mass to intrinsic luminosity is constant among galaxies.  Therefore,
\begin{equation}
mass \propto 10^{-M_{b(j)} / 2.5}
\label{eq:0},
\end{equation}
where $M_{b(j)}$ (in units of magnitude) is the intrinsic absolute B band magnitude $M_b$ of the $j^{th}$ neighbor galaxy.  The adjective ``intrinsic'' means after the luminosity effects of neighbor galaxies have been subtracted (Paper 1).

The force $F_{(k)}$ on a particle in the $k^{th}$ target galaxy exerted by a number $n$ of neighboring close galaxies is,
\begin{equation}
\bm F_{(k)} = \sum^n_{j=1} G_m m 10^{-M_{b(j)} / 2.5} \frac{\bm R_{(jk)}}{R^3_{(jk)}}
\label{eq:1},
\end{equation}
where $\bm R_{(jk)}$ is the vector distance from the $k^{th}$ target galaxy to the $j^{th}$ neighbor galaxy, $G_m$ is the constant of proportionality such that the units of $F_{(k)}$ are dynes, and $m$ is the mass of the test particle in the target galaxy.  The bold letters denote vectors.  The same letter not bold denotes the magnitude of the vector.  The letters in parentheses are indices denoting the galaxy.  For ease of calculation $M_{b(j)}$ is without the correction found in Paper 1.  Hence, the $G_m$ factor also contains a constant which is the average change in luminosity of the $j^{th}$ galaxy due to re-emitted radiant energy from neighboring galaxies.  From Paper 1, the deviation from average of the re-emitted energy will produce an error in $ M_{b(j)}$ of $\pm$ 0.3 magnitude.

Define $\bm F_{m(k)}$ as $\bm F_{(k)}$ in a coordinate system that is galactocentric, right-handed, and Cartesian,
\begin{equation}
\bm F_{m(k)} \equiv \frac{1}{G_m m} ( \bm F_{minor(k)} + \bm F_{major(k)} + \bm F_{polar (k)})
\label{eq:2},
\end{equation}
where $\bm F_{minor(k)}$, $\bm F_{major(k)}$, and $\bm F_{polar (k)}$ are the component vectors of $\bm F_{(k)}$ in the direction of the minor axis, major axis, and polar axis of the $k^{th}$ target galaxy, respectively.  Since the $\bm F_{major(k)}$ acts radially along the major axis, the effect of the central force of a galaxy $\bm F_c$ along the major axis is reduced if the measurement of asymmetry is at the same radius from the center on each side of the galaxy.  Define the rotation curve asymmetry $v_d$ as the maximum difference in rotation velocity along the major axis between one side of the center of a galaxy to the other measured at the same radius from the center of the galaxy.

Consider the effect of $\bm F_{m(k)}$ on $v_d$ is exerted proportionally by a vector $\bm K$,
\begin{equation}
v_d = \vert \bm K \, \bullet \, \bm F_{m(k)} \vert
\label{eq:3},
\end{equation}
\begin{equation}
\bm K = K_{minor} \bm e_{minor} + K_{major} \bm e_{major} + K_{polar} \bm e_{polar} 
\label{eq:4},
\end{equation}
where $\bm e_{minor}$, $\bm e_{major}$, and $\bm e_{polar}$ are the unit vectors in the minor axis direction, the major axis direction, and the polar axis direction of the $k^{th}$ target galaxy, respectively, $K_{minor}$, $K_{major}$, and $K_{polar}$ are coefficients of the corresponding unit vectors, and the $\vert \, \, \vert$ means the absolute value.  The direction of $\bm e_{major} $ is directed eastward along the axis defined by the position angle (PA).  The $\bm e_{minor} $ is perpendicular to $\bm e_{major} $, in the plane of the galaxy, and directed away from the Milky Way.  A right-handed coordinate system means $\bm e_{minor} \times \bm e_{major} = \bm e_{polar}$.  In conformity to Paper 1, a positive orientation sign implies $\bm e_{minor} $ has a component directed southward and $\bm e_{polar}$ has a component directed away from the Milky Way.  A negative orientation sign implies $\bm e_{minor} $ has a component directed northward and $\bm e_{polar}$ has a component directed toward the Milky Way. 

The kinematic minor axis is the direction along which the line-of-sight motion equals the systemic velocity.  The kinematic major axis is the direction along which the line-of-sight motion peaks as a function of angle about the galactic center.  The tilted ring model predicts the kinematic minor axis is perpendicular to kinematic major axis.  Galaxies with warped disks may have curved kinematic axes.  In this Paper, the major and minor axes are not kinematic axes when a warp is present.

The $\bm K$ is found by plotting $v_d$ versus $\vert \bm K \bullet \bm F_{m(k)} \vert$ such that a best fit of equation~(\ref{eq:3}) results.

Since the distance to neighboring galaxies is much greater than the diameter of galaxies, the first approximation is that $\bm F_{m(k)}$ is uniform across a galaxy.  The asymmetry in the HI rotation curve is a sensitive tracer of $\vert \bm K \bullet \bm F_{m(k)} \vert$.

\section{DATA AND ANALYSIS}

The criteria for choosing target galaxies for the analysis are (1) the galaxy has a published Cepheid distance $D_c$ and a published HI rotation curve asymmetry, (2) the distance to the galaxy must be large enough that the Milky Way's contribution is negligible, (3) the distance to the galaxy must be large enough that the contribution of peculiar velocity to the redshift $z$ measurement is relatively small among the galaxies around the target galaxy, and (4) a published apparent magnitude in the B band $m_b$ must be available for both the target and close galaxies.  Nine galaxies were found to meet these criteria.

The data for the sample of nine target galaxies are presented in Table~\ref{tab:1}.  The morphology was taken from the NED database\footnote{The NED database is available at: http://nedwww.ipac.caltech.edu.}.  The rotation curve shape and measured $v_d$ were taken from the references.  The galaxy orientation sign which was used to determine the component vectors of $\bm F_{m(k)}$ was taken from Paper 1.  The total apparent corrected B-band magnitude $m_b$ is ``btc'' in the LEDA database \footnote{The LEDA database is available at: http://leda.univ-lyon1.fr. }.  The value of ``btc'' for close galaxies was used since I band data is generally unavailable.  The $D_c$ was taken from \citet{free} except as noted in Table~\ref{tab:1}.

The sample of target galaxies has low surface brightness (LSB), medium surface brightness (MSB), and high surface brightness (HSB) galaxies, galaxies with a range of 21-cm line width at 20\% of peak value corrected for inclination $W^i_{20}$ from 122 $km s^{-1}$ to 654 $km s^{-1}$, includes LINER, Sy, HII and less active galaxies, galaxies with a absolute B magnitude (``mabs'' from the LEDA database) from -22.11 mag. to -17.03 mag., a distance range from 1.13 Mpc to 14.73 Mpc, and field and cluster galaxies.

The NED database was used to assemble a list of galaxies closest to each of the target galaxies.  The selection of close galaxies used the Hubble Law with a Hubble constant $H_o$ of 70 $km s^{-1}Mpc^{-1}$ and galactic z to calculate the distances to galaxies.  The angular coordinates were taken from the NED database.  For each target galaxy, the distance from each target galaxy to each of the other galaxies was calculated.  The neighbor galaxies with the smallest distance were chosen as the close galaxies.

The uncertainty of the value of $H_o$ is high.  The use of $H_o$ herein is restricted to only choosing the close galaxies to each target galaxy.  As the relative distances become larger, the distance uncertainty increases if only because of a greater variation in peculiar velocity.  Therefore, there is a practical upper limit on the number of galaxies that can be used to evaluate $\bm F_{m(k)}$.  If too few close galaxies are used, $\bm F_{m(k)}$ will be underestimated.  Following Paper 1, the number of close galaxies found to produce the highest correlation coefficient was 10.

The rotation curve of NGC 4321 at 4 arcmin implies it has a declining shape.  NGC 4321 has a large asymmetry in the south west lobe of 209 $km \, s^{-1}$ which cannot be interpreted as disk rotation \citep{knap}.  The H$\alpha$ rotation curve is flat out to 20 kpc with $v_d = 52.4 km \, s^{-1}$ \citep{arse}.  The receding side of the HI rotation curve of NGC 4321 is declining and appears more normal.  The approaching side of the HI rotation curve of NGC 4321 is ``lopsided'' \citep{knap} and is sharply rising after a slight decline.  \citet{guha} suggested the HI rotation curve is declining.

The $R_{( jk)}$ and $ M_{b(j)} $ in equation (\ref{eq:1}) were calculated as in Paper 1.  Because $ M_{b(j)} $ is more uncertain without a secure distance ($D_c$) to each galaxy, the value used for $ M_{b(j)} $ was not the intrinsic magnitude.  Therefore, the value of $\bm F_{m(k)}$ was overestimated.  This causes the value of $\vert \bm K \bullet \bm F_{m(k)} \vert $ to have a larger scatter.

\begin{figure}
\plotone{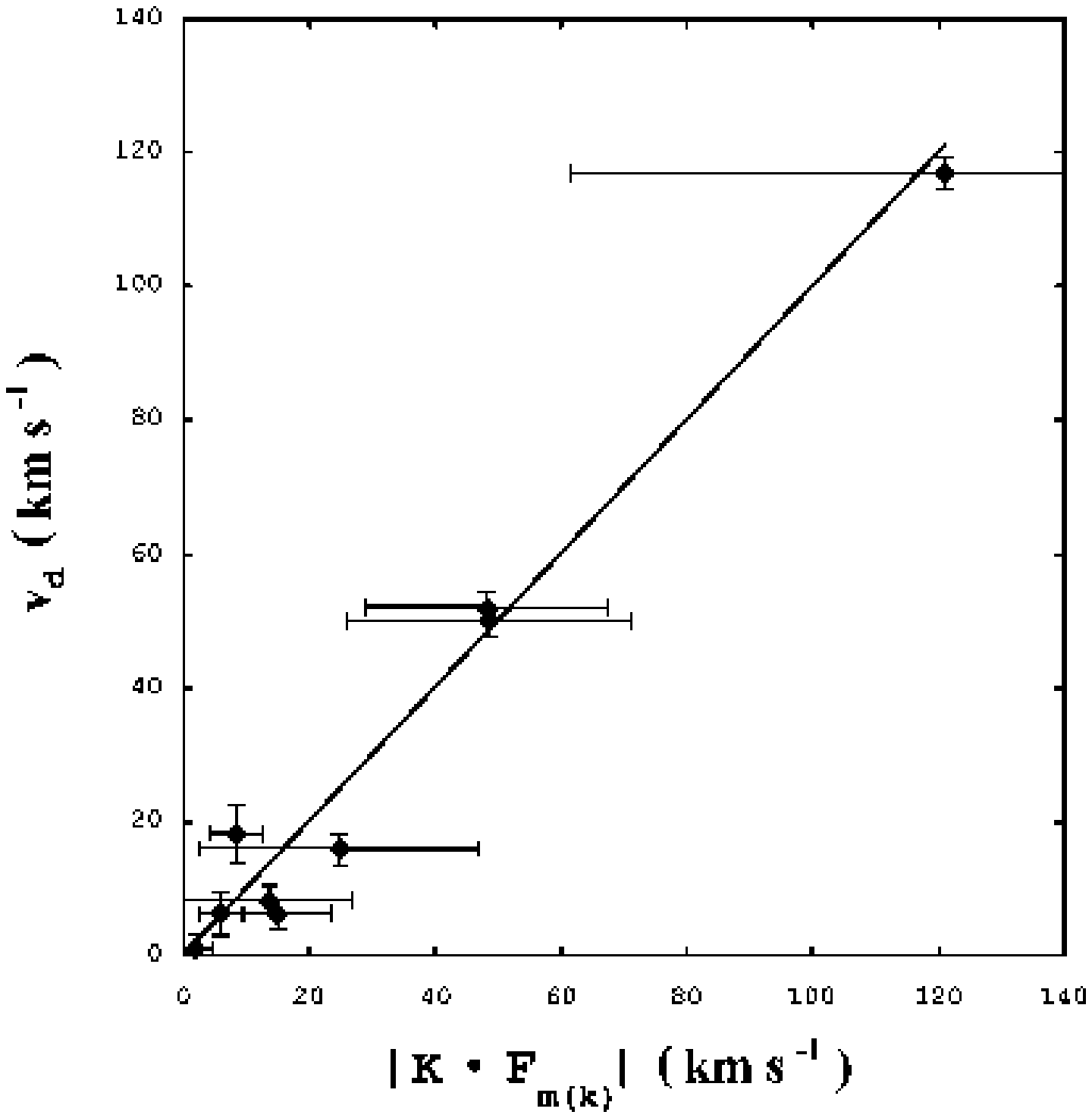}
\caption{\label{fig:1} Plot of $v_d$ $( km \, s^{-1} )$ versus $\vert \bm K \bullet \bm F_{m(k)} \vert$ $( km \, s^{-1} )$.  The equation of the line is  $v_d = \vert \bm K \bullet \bm F_{m(k)} \vert$ where $\bm K = ( 9.70 \times 10^{-7} ) \bm e_{minor} + ( 1.88 \times 10^{-7} ) \bm e_{major} - (4.24 \times 10^{-8}) \bm e_{polar} \, \,km \, s^{-1}\,dyne^{-1} G_m m $.  The line is a best fit of the data.}
\end{figure}

Figure~\ref{fig:1} is a plot of  $v_d$ versus $\vert \bm K \bullet \bm F_{m(k)} \vert $.  The coefficients of $\bm K$ were determined such that variance between the line with a slope of one and data (F test) is maximized and the sum of the difference between the points and the line is minimized, 
\begin{equation}
v_d = S \, \vert \bm K \bullet \bm F_{m(k)} \vert \, + I
\label{eq:5},
\end{equation}
where 
\begin{eqnarray}
\bm K & = & ( 9.70 \times 10^{-7} ) \bm e_{minor} + (1.88 \times 10^{-7} ) \bm e_{major} \nonumber \\*
& & - (4.24 \times 10^{-8}) \bm e_{polar} \nonumber \\*
& & \, \,km \, s^{-1}\,dyne^{-1} G_m m 
\label{eq:6},
\end{eqnarray}
$K = 9.89 \times 10^{-7}  \, \,km \, s^{-1}\,dyne^{-1} G_m m $, the slope $S = 1.00 \pm 0.06 $, and the intercept $I = 0 \pm 3 km\, s^{-1}$.  The error is 1 $\sigma$, the correlation coefficient is 0.99, and the F test is 0.99.

\begin{figure}
\plotone{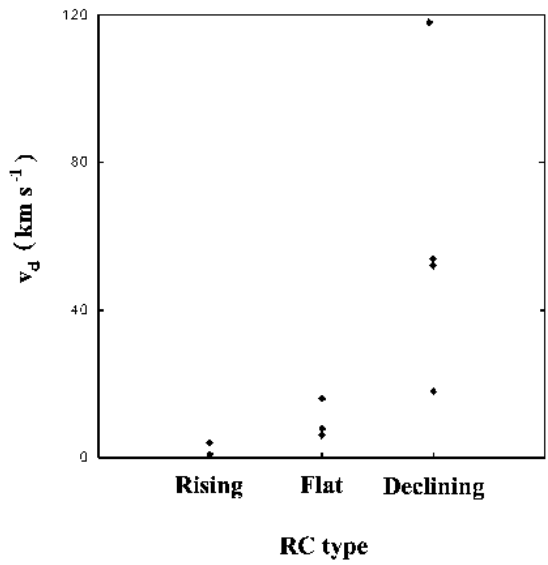}
\caption{\label{fig:2} Plot of $v_d$ $( km \, s^{-1} )$ versus rotation curve type.  NGC 4321 has $v_d = 117 \, km \, s^{-1} $.  It's shape is either flat \citep{arse} or declining \citep{guha}.}
\end{figure}

Figure~\ref{fig:2} is a plot of $v_d$ versus rotation curve type.  Figure~\ref{fig:2} suggests a qualitative correlation between  $v_d$ and the slope of the sample galaxies' rotation curve in the disk region.

\section{DISCUSSION }

The $v_d$ is defined at the radius from the center along the major axis with the maximum asymmetry in rotation velocity.  We speculate this implies the dynamics of galaxies are a function of radius from the center rather than a number of scaling factors.

Since both Paper 1 and this Paper require the orientation sign of the target galaxy to be established, the method of this Paper may confirm the orientation sign that the Paper 1 method finds.

Paper 1 used luminosity ($M_b$) as an indicator of the radiant energy from a neighbor galaxy.  This Paper used luminosity as an indicator of mass in a neighbor galaxy.  Luminosity, mass, and radiant energy of a galaxy appear tightly related.  The difference between radiant energy's action and potential energy's action on a target galaxy is in the way energy is transferred.  The amount of radiant energy transferred to a target galaxy depends on the surface area presented to the neighbor galaxy.  Hence, the radiant energy transferred is maximal if the neighbor galaxy is along the target galaxy's polar axis and minimal if the neighbor galaxy is in the plane of the target galaxy.  The potential energy transferred is at a maximum if the neighbor galaxy is in the target galaxy's plane and at a minimum if the neighbor galaxy is along the polar axis of the target galaxy.

Paper 1 concluded that the contribution of the conservative, potential field of neighboring galaxies contributed no energy to the B-band luminosity because matter in a galaxy orbited in radial shells of constant mass.  The target galaxy's movement in the intergalactic potential field could cause only minimal heat change.  The correlation coefficient between the Paper 1 correction factor $C_f$ and $v_d$ is -0.04, between $C_f$ and $\vert \bm K \bullet \bm F_{m(k)} \vert$ is -0.09, between the Paper 1 impinging energy $E_a$ and $v_d$ is 0.61, and between $E_a$ and $\vert \bm K \bullet \bm F_{m(k)} \vert$ is 0.65.  These poor correlations imply the potential energy field of the close galaxies has little or no effect on the $M_b$ measurement of the target galaxies.

\citet{pers,pers2,sofu2} found a strong correlation between a galaxy's luminosity and rotation curve shape.  Low luminosity galaxies have rising rotation curves; high luminosity galaxies have flat and declining rotation curves.  This Paper found close neighbor galaxies appear to cause higher asymmetry in the rotation curve and to cause flat and declining rotation curves.  Also, this paper suggests that without the influence of neighboring galaxies (intrinsic), the rotation curve would be rising.  Paper 1 found close neighbor galaxies appear to cause higher luminosity in a galaxy depending on the orientation of the galaxy relative to the close galaxies.  Therefore, neighboring galaxies cause differences in the mass/luminosity ratio among galaxies by increasing the apparent luminosity as outlined in Paper 1.

The dark matter model (DMM) expects the intrinsic rotation curve to be declining.  If dark matter (DM) is considered part of a galaxy's mass, the constant intrinsic mass to luminosity ratio among galaxies implies the ratio of DM to luminous mass is also constant among galaxies.  To have a rising rotation curve caused by DM implies the potential caused by the DM must increase with radius while the luminous mass is decreasing with radius.  If galaxies have differing mass because of differing mass infall rates, then the constant ratio of DM to luminous mass and the required DM distribution seems implausible.

The linear potential model (LPM) explored by Mannheim and Kazanas \citep{mann} proposes a gravitational potential of the form $V(r) = - \beta / r +\gamma r / 2$.  The LPM predicts rising rotation curves are the intrinsic shape and flat rotation curves are a result of mass distribution within the galaxy.  Also, the LPT predicts declining rotation curves require other galaxy influence with a linear potential and a very delicate balance.  The modified Newtonian dynamics (MOND) model \citep{bott} expects the intrinsic rotation curve to be flat and flat rotation curves to be axis symmetric.    The finding of this Paper is inconsistent with the LPM, the MOND model, and the DMM.

\section{CONCLUSION}

The rotation curve asymmetry is linearly related to the force from neighboring galaxies.  The relation is  $v_d = 1.00(\pm 0.06) \, \vert \bm K \bullet \bm F_{m(k)} \vert \, + 0(\pm3)$ at 1 $\sigma$ where $\bm K = ( 9.70 \times 10^{-7} ) \bm e_{minor} + (1.88 \times 10^{-7} ) \bm e_{major} - (4.24 \times 10^{-8}) \bm e_{polar} \, \, km \, s^{-1}\,dyne^{-1} G_m m$.  The correlation coefficient is 0.99 with F test of 0.99.

Rising rotation curves have little rotation curve asymmetry and flat and declining rotation curves have greater asymmetry.

Paper 1 is correct in assuming the potential field of neighboring galaxies has no measurable impact on the $M_b$ of target galaxies.

The orientation sign adopted for Paper 1 is confirmed for the galaxies in this Paper.  This includes the problematic galaxies in Paper 1 NGC 3319 and NGC 4535.

\acknowledgments
This research has made use of the NASA/IPAC Extragalactic Database (NED), which is operated by the Jet Propulsion Laboratory, California Institute of Technology, under contract with the National Aeronautics and Space Administration.

This research has made use of the LEDA database (http://leda.univ-lyon1.fr).

We acknowledge and appreciate the financial support of Cameron Hodge, Stanley, New York, and Maynard Clark, Apollo Beach, Florida, while working on this project.

\clearpage

\clearpage

\clearpage

\begin{deluxetable}{llcrcrcr}
\tabletypesize{\scriptsize}
\tablecaption{\label{tab:1} Data for the chosen galaxies.}
\tablehead{\colhead{Galaxy} &\colhead{NED type \tablenotemark{a}} &\colhead{sign \tablenotemark{b}} &\colhead{$D_c$ \tablenotemark{c}} &\colhead{RC \tablenotemark{d}} &\colhead{$v_d$ \tablenotemark{e}} &\colhead{ref \tablenotemark{f}} &\colhead{$\vert \bm K \bullet \bm F_{m(k)} \vert$ \tablenotemark{g}} }
\startdata
\objectname[]{NGC 0925} &SAB(s)d;HII&-&9.13$\pm$0.17&D&52$\pm$2&1&48$\pm$20\\
\objectname[]{NGC 2403} &SAB(s)cd&-&3.14$\pm$0.36&F&8$\pm$2&2&14$\pm$10\\
\objectname[]{NGC 2841} &SA( r)b;LINER Sy&+&14.07$\pm$1.57\tablenotemark{h}&D&16$\pm$2&3&25$\pm$10\\
\objectname[]{NGC 3031} &SA(s)ab;LINER Sy1.8&-&3.55$\pm$0.09&D&18$\pm$4&4&8$\pm$\phm{0}4\\
\objectname[]{NGC 3109} &SB(s)m&+&1.13$\pm$0.12\tablenotemark{i}&R&1$\pm$2&5&2$\pm$\phm{0}2\\
\objectname[]{NGC 3198} &SB(rs)c&-&13.69$\pm$0.50&F&6$\pm$3&6&6$\pm$\phm{0}3\\
\objectname[]{NGC 3319} &SB(rs)cd;HII&-&13.44$\pm$0.57&R&6$\pm$2&7&15$\pm$\phm{0}8\\
\objectname[]{NGC 4321} &SAB(s)bc;LINER HII&-&14.33$\pm$0.46&D&117$\pm$2&8&121$\pm$60\\
\objectname[]{NGC 4535} &SAB(s)c&-&14.80$\pm$0.32&D&50$\pm$2&9&49$\pm$20\\
\enddata

\tablenotetext{a}{Galaxy morphological type from the NED database.}
\tablenotetext{b}{The sign is the orientation of the polar axis of the target galaxy.  A ``+'' sign indicates the polar axis is directed away from the Milky Way.  The sign is taken from Paper 1.}
\tablenotetext{c}{ Distance in Mpc from Cepheid data from \citet{free}, unless otherwise noted. }
\tablenotetext{d}{ Galaxy's HI rotation curve type according to slope in the outer disk region.  R is rising, F is flat, and D is declining. }
\tablenotetext{e}{The maximum difference in rotation velocity ($km s^{-1}$) in the disk region along the major axis from one side of the galaxy to the other at the same radius.}
\tablenotetext{f}{References are for the rotation curve data for each galaxy.}
\tablenotetext{g}{In units of $km s^{-1} $.}
\tablenotetext{h}{Distance is from \citet{macr}.}
\tablenotetext{i}{Distance is from \citet{ferr}.}
\tablerefs{
(1) \citet{krum}; (2) \citet{shos}; (3) \citet{fill}; (4) \citet{gott2}; (5) \citet{jobi}; (6) \citet{vana}; (7) \citet{moor}; (8) \citet{knap}; (9) \citet{chin}}

\end{deluxetable}

\clearpage

\end{document}